\documentclass[apj]{emulateapj}
\usepackage{comment}
\usepackage{ifthen}
\usepackage{amsmath}
\usepackage{amsfonts}
\usepackage{amssymb}
\usepackage{multirow}
\usepackage{subfigure}
\usepackage{epsfig}

\usepackage[normalem]{ulem}
\usepackage{color}

\usepackage[utf8]{inputenc}
\usepackage{graphicx}

\newcommand{\logg}{$\log g$}
\newcommand{\teff}{$T_{\rm eff}$}
\newcommand{\msun}{M$_\odot$}
\newcommand{\fl}{$F_8$}

\begin{document}

\title{A Granulation ``Flicker''-based Measure of Stellar Surface Gravity}
\author{Fabienne A. Bastien\altaffilmark{1,2}, Keivan G. Stassun\altaffilmark{3,4}, Gibor Basri\altaffilmark{5}, Joshua Pepper\altaffilmark{6,3}}

\altaffiltext{1}{Department of Astronomy and Astrophysics, 525 Davey Lab, The Pennsylvania State University, University Park, PA 16803}
\altaffiltext{2}{Hubble Fellow}
\altaffiltext{3}{Vanderbilt University, Physics \& Astronomy Department, 1807 Station B, Nashville, TN 37235, USA}
\altaffiltext{4}{Fisk University, Department of Physics, 1000 17th Ave. N, Nashville, TN 37208, USA}
\altaffiltext{5}{Astronomy Department, University of California, Berkeley CA 94720}
\altaffiltext{6}{Department of Physics, Lehigh University, 16 Memorial Drive East, Bethlehem, PA, 18015}

\begin{abstract}
In \citet{bastien13} we found that high quality light curves, such as those obtained by {\it Kepler}, may be used to measure stellar surface gravity via granulation-driven light curve ``flicker'' (\fl).  Here, we update and extend the relation originally presented in \citet{bastien13} after calibrating \fl\ against a more robust set of asteroseismically derived surface gravities.  We describe in detail how we extract the \fl\ signal from the light curves, including how we treat phenomena, such as exoplanet transits and shot noise, that adversely affect the measurement of \fl.  We examine the limitations of the technique, and, as a result, we now provide an updated treatment of the \fl-based \logg\ error.  We briefly highlight further applications of the technique, such as astrodensity profiling or its use in other types of stars with convective outer layers.  We discuss potential uses in current and upcoming space-based photometric missions.  Finally, we supply \fl-based \logg\ values, and their uncertainties, for 27\,628 {\it Kepler} stars not identified as transiting-planet hosts, with 4500$<$\teff$<$7150 K, 2.5$<$\logg$<$4.6, $K_p \le$13.5, and overall photometric amplitudes $<$10 parts per thousand.
\end{abstract}

\section{Introduction}

NASA's {\it Kepler} mission simultaneously observed over 150\,000 Sun-like stars in the constellation Cygnus for more than four years.  Its primary aim was to determine the occurrence rate of Earth-like planets in the habitable zones of such stars, and it was therefore designed to achieve milli- to micro-magnitude photometric precision, thereby pushing astrophysical studies into regimes inaccessible from the ground.  Indeed, high precision photometric surveys like NASA's {\it Kepler} have generated much excitement through the discoveries they have enabled, from the discovery of thousands of extrasolar planetary candidates \citep{borucki11,batalha13,burke14} in a wide variety of orbital configurations \citep{sanchis-ojeda13} and environments \citep{meibom13} and whose host stars span a range of spectral types (from A \citep{szabo11} to M \citep{muirhead12}) to short timescale variability in active galactic nuclei \citep{mushotzky11,wehrle13}.  In particular, these space-based missions have revitalized stellar astronomy, permitting the first ensemble asteroseismic analyses of Sun-like stars \citep{chaplin11a,huber11,stello13}, as well as rotation period \citep{reinhold13,nielsen13,walkowicz13,mcquillan14}, flare \citep{walkowicz11,maehara12,notsu13} and differential rotation \citep{reinhold13,lanza14,aigrain15} studies of very large numbers of field stars, among others.  In addition to revealing a wide variety of stellar variability \citep{basri10}, the nearly uninterrupted coverage has enabled different and new characterizations of this variability \citep{basri11}, including several different proxies for chromospheric activity \citep{chaplin11b,mathur14}.

Unfortunately, both the fulfillment of the key mission objective and the large ensemble studies of the variability of Sun-like stars suffer significantly from our poor knowledge of the fundamental stellar parameters, specifically the stellar surface gravity (\logg).  The mission primarily aimed to monitor dwarf stars, and the sheer number of potential targets in the {\it Kepler} field required an efficient observing strategy to weed out as many likely evolved stars as possible \citep{brown11}.  The expectation was that the wider community would perform extensive follow-up observations and analyses in order to refine the stellar parameters of the {\it Kepler} stars, in particular the planet hosts.  Extensive spectroscopic campaigns, largely focused on the planet host stars, have contributed improved effective temperatures, metallicities and \logg\ of many hundreds of stars \citep{buchhave12,buchhave14,marcy14,petigura13}, accompanied by key new insights into the nature and diversity of exoplanetary systems.  Asteroseismology has yielded highly precise stellar gravities and densities for hundreds of dwarf and subgiant stars \citep[e.g.,][]{chaplin14} and thousands of giant stars \citep[e.g.,][]{stello13}.  Methods that rely on transiting bodies (eclipsing binary stars and transiting exoplanets) have also yielded improved stellar parameters \citep{conroy14,parvizi14,plavchan14}, albeit on a more limited number of targets.

Outside of asteroseismology, less effort has gone into improving the stellar parameters of the wider {\it Kepler} sample, largely because the transiting planet host stars alone are already severely taxing ground-based follow-up resources; indeed, many of the planet hosts still have not been fully characterized to date.  Yet both extrasolar planet studies (specifically planet occurrence analyses) and ensemble analyses of stellar variability require knowledge of the properties of the stars that do not host transiting exoplanets.  Techniques and analyses that can yield more accurate stellar parameters for the larger {\it Kepler} sample, in particular at relatively low (resource) cost, can therefore be very useful to both the stellar astrophysics and extrasolar planet communities.

In \citet{bastien13}, we found that the relatively high frequency stellar variations observed by {\it Kepler} --- those occurring on less than 8 hour timescales and which we dub ``flicker'' (\fl) --- do indeed encode a simple measure of fundamental stellar parameters.  This work, in addition to introducing a stellar evolutionary diagram constructed solely with three different characterizations of light curve variability, demonstrated that the granulation-driven \fl\ can yield the stellar surface gravity with a precision of $\sim$0.1--0.2 dex.  This builds on previous works that have demonstrated that granulation is imprinted in asteroseismic signals and that this signal correlates strongly with stellar surface gravity \citep{kjeldsen11,mathur11}.

On the one hand, \fl\ permits the measurement of fundamental stellar properties directly from high-precision light curves, which can in turn facilitate the determination of the properties of extrasolar planets \citep{kipping14,bastien14b} and possibly shed light on the nature of the radial velocity jitter that impedes planet detection \citep{cegla14,bastien14a}.  On the other hand, this work can improve our understanding of stellar structure and evolution by, for example, enabling us to place observational constraints on granulation models \citep{cranmer14,kallinger14}.

Here, we expand upon the results presented in \citet{bastien13} by detailing the steps used to measure \fl\ (Section \ref{sec:analysis}), outlining the limitations of and constraints on the method as currently defined while updating the relations presented in \citet{bastien13} (Section~\ref{sec:method}), providing \fl-based \logg\ values for 27\,628 {\it Kepler} stars (Section~\ref{sec:results}), and briefly exploring the theory behind and some applications of \fl\ (Section~\ref{sec:discussion}) before concluding in Section~\ref{sec:conclusion}.

\section{Data Analysis}\label{sec:analysis}

We begin this section by describing some of the characteristics of the {\it Kepler} data.  We follow this by outlining the steps we take to measure \fl\ in the {\it Kepler} light curves, expanding upon the level of detail previously provided in \citet{bastien13}.  Finally, we describe the asteroseismic calibration datasets we use to place the \fl-\logg\ scale on an externally validated absolute scale --- updated from the calibration dataset used in \citet{bastien13} --- and to assess the precision and accuracy of our \fl-\logg\ measurements.

\subsection{{\it Kepler} data}

Previous studies \citep[such as][]{bor10,koch10,jenk10a,jenk10b}, describe the {\it Kepler} mission data products in detail.  Here we provide a brief summary of the data as relevant for our analysis.

In total, the {\it Kepler} mission observed over 200~000 stars, with $\sim$160~000 observed at any given time.  The vast majority of stars were observed in long cadence (29.4~min. co-adds), and $\sim$512 stars were observed in short cadence (58.8~sec. co-adds).  The {\it Kepler} spacecraft's orbital period is 371 days, so once every 93 days, the spacecraft is rotated 90 degrees to re-orient its solar panels.  Each of these rolls represents a division between epochs of data, and so the data are organized in so-called Quarters, although not every quarter is precisely 1/4 of a solar year.  Specifically, Q0, the commissioning data, is only 9.7 days long, and Q1, the first science quarter, is 33.5 days long.  Subsequent quarters are all approximately 90 days long. 

The data from the {\it Kepler} mission contain artifacts and systematic features unique to the {\it Kepler} telescope.  For full details on these issues, see the {\it Kepler} Instrument Handbook, the {\it Kepler} Archive Manual and the latest version of the Data Release Notes \citep{thompson15} at MAST. Of particular note, the data available at MAST contain Simple Aperture Photometry (SAP), as well as Pre-search Data Conditioning, Maximum A Posteriori \citep[PDC-MAP;][]{stumpe14} versions of the {\it Kepler} light curves.  The goal in producing the PDC-MAP versions of the light curves is to remove all possible behavior in the light curves that could interfere with the detection of transiting exoplanets.  In general, real astrophysical signals on timescales shorter than about 20 days are preserved; on longer timescales, they might be removed \citep{stumpe14}. In our analysis we use all quarters except for Q0, and we only use the long cadence light curves.  Additionally, we only use the PDC-MAP light curves, as further discussed in Section~\ref{sec:sap_vs_pdc}.

\subsection{Measuring Flicker\label{sec:measf8}}

The flicker method is at present based on the use of {\it Kepler} long-cadence (30-min sampled) light curves, with the standard pipeline produced fluxes (PDC-MAP).  The steps involved in measuring the flicker amplitude in a {\it Kepler} light curve, described below, include: (1) clipping of outliers and removal of known transit events, (2) smoothing on multiple timescales to isolate the 8-hr flicker signal of interest, (3) removal of the instrumental (i.e., non-astrophysical) ``flicker" due to detector shot noise, and (4) incorporating knowledge of quarter-specific aperture contamination to mitigate effects of pointing jitter, etc.  As further described in section~\ref{sec:fluxfrac}, for each star, we measure \fl\ from each available quarter of data and take the median or robust mean of the measurements as our final measure of \fl.

\subsubsection{Sigma clipping and transit removal\label{sigclip}}

The photometric flicker is fundamentally a measure of the  r.m.s.\ of the light curve on timescales shorter than some maximum timescale (8-hr in our current implementation) caused by astrophysical ``noise" in the integrated stellar flux. Therefore as a first step it is essential to eliminate spurious data outliers in the light curve that arise from either non-astrophysical effects (e.g., data glitches) or from punctuated astrophysical effects that are unrelated to the surface granulation that drives the fundamental \fl-\logg\ relation (e.g., flares, transits). In all that follows, we perform a simple linear interpolation of the light curve across any data gaps (flagged by the {\it Kepler} pipeline as {\tt NaN} values) in order to preserve the intrinsic timescales in the original light curve data sampling.

We then sigma-clip the light curve to remove both random individual data outliers and random short-duration strings of data points arising from impulsive flares. We do this via an iterative 2.5-$\sigma$ clip, in which we flag as {\tt NaN} individual data points that deviate by more than 2.5 times the r.m.s.\ of the full light curve, and we iterate the clip until no additional data points are removed.  We employ the 2.5-$\sigma$ cut in order to remove the 1\% most outlying points.

We found that some large-amplitude photometric variations are not adequately removed by the simple smoothing that we apply to isolate the short-timescale flicker variations (see Sec.\ \ref{sec:smooth}). Therefore, prior to sigma-clipping we first subtract a low-order quadratic spline fit to the light curve.  This spline subtraction serves to remove long-timescale fluctuations such as rotationally modulated variations due to spots, pulsations, long-duration stellar eclipses, etc., while preserving the short-timescale variations that we are ultimately interested in measuring via flicker. 

Known transits due to putative planetary bodies also inflate the r.m.s.\ of the light curve, but these are usually of sufficiently short duration and/or of sufficiently shallow depth that they may be missed by the low-order spline subtraction and/or by the sigma-clipping. Therefore we can also mask out as {\tt NaN} any data points within $\pm$0.025 phase of the transits tabulated in the NASA Exoplanet Archive prior to the sigma-clipping. Note that in the present study we do not include stars known to host transiting planets; instead see \citet{bastien14b} for an analysis of \fl-based \logg\ for the bright KOIs.

\subsubsection{Smoothing\label{sec:smooth}}

Following \citet{basri11} and \citet{bastien13}, we determine the light curve flicker on an 8-hr timescale by first subtracting a smoothed version of the light curve from itself, where the smoothing timescale is 8-hr (i.e., 16 long-cadence time bins) and then measuring the r.m.s.\ of the residual light curve. In our current implementation, the smoothing is done with a simple 16-point boxcar using the {\tt smooth} function in {\tt idl}. We treat gaps in the light curve as {\tt NaN} in order to preserve the true data cadence when performing the 8-hr smooth and r.m.s.\ calculations.

\subsubsection{Removal of shot noise contribution to flicker\label{sec:f8shotnoise}}

The flicker measure we seek should represent the true astrophysical noise arising from the stellar surface granulation, with significant contributions from acoustic oscillations for evolved stars (up to $\sim$30\% for red giants; see \citet{kallinger14}), and so it is necessary to remove the contribution of shot noise to the observed flicker signal. In \citet{bastien13}, we used the full set of stars observed by {\it Kepler} to define a quadratic fit to the bottom 0.5\%-ile of 8-hr r.m.s.\ versus apparent magnitude, representing the empirical shot noise floor of the {\it Kepler} data as a function of magnitude. We then subtracted this shot noise in quadrature from the observed r.m.s.\ for a given target as appropriate for its {\it Kepler} apparent magnitude.  We note that a very small fraction of stars will have negative \fl\ values resulting from over-subtracted shot noise. In such cases, we assume that the stars have the smallest possible \fl\ (i.e., that they have the highest possible \logg\ and must lie on the main sequence), and we accordingly assign a \logg\ that corresponds to the main sequence value as follows:
\begin{equation*}
\log g = \begin{cases}
4.6 & T_{\rm eff} < 5000\, {\rm K} \\
4.5 & 5000\, {\rm K} \leq T_{\rm eff} < 6000\, {\rm K} \\
4.35 & T_{\rm eff} \geq 6000\, {\rm K}
\end{cases}
\end{equation*}

\begin{figure}[ht]
\includegraphics[width=8.5 cm]{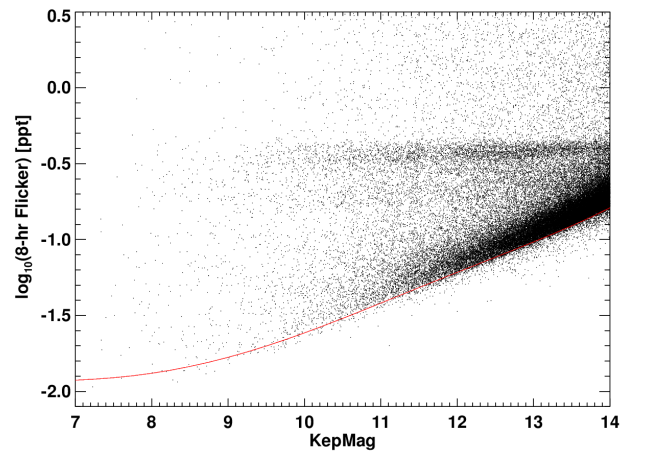}  
\caption{\label{fig:kpcorr}
{\it Removal of Shot Noise Contribution to \fl.}  Plotting \fl\ as a function of Kepler magnitude shows the increasing importance of shot noise to the measured flicker signal.  We correct our \fl\ measurements by fitting the lower envelope of the shown distribution (red curve) and subtracting this fit from the measurement.  See Section~\ref{sec:f8shotnoise}}
\end{figure}

The quadratic fit from \citet{bastien13} was optimized for the stars brighter than $\sim$12th mag that were the primary focus of that study, but we found that this under-estimates the shot noise for fainter stars. We have therefore extended the empirical shot-noise fit from \citet{bastien13} by adding a second, higher order polynomial, to better capture the residual shot noise as a function of magnitude for stars in the magnitude range 12--14. We show the updated two-component polynomial fit in Figure~\ref{fig:kpcorr}.  Our final set of {\it Kepler} magnitude relations is:
\begin{equation}
\begin{split}
min_{1}(\log_{10}F_8)= -0.03910 - 0.67187K_p \\ + 0.06839K^{2}_{p} - 0.001755K^{3}_{p}
\end{split}
\end{equation}
\begin{equation}
\begin{split}
min_{2}(\log_{10}F_8)= -56.68072 + 29.62420K_p \\ - 6.30070K^{2}_{p} + 0.65329K^{3}_{p} \\ - 0.03298K^{4}_{p} + 0.00065K^{5}_{p}
\end{split}
\end{equation}

Note that these polynomial fits are performed to the logarithmic $F_8$ values. In addition, the fits are defined such that the shot-noise correction to the observed flicker is done in quadrature, i.e., 
\begin{equation}
F_{8,corr} = \left[ F_{8,obs}^2 - (10^{min_1})^2 - (10^{min_2})^2 \right]^{1/2}
\label{eq:f8corr}
\end{equation}

\subsubsection{Flux fraction and neighbor contamination\label{sec:fluxfrac}}

The {\it Kepler} spacecraft is known to experience small offsets in the precise pixel positions of stars from quarter to quarter as a result of the quarterly spacecraft ``rolls". These small pixel offsets result in small but measureable changes in (a) the amount of a star's light that is included in its pre-defined aperture and (b) the amount of light from neighboring stars' entering the target star's aperture. As a result, in some cases a given star's photometry for one or more quarters may include an unacceptably large contamination from neighboring stars, resulting in a larger observed \fl\ for those quarters. 

\begin{figure}[ht]
\includegraphics[width=0.45\linewidth]{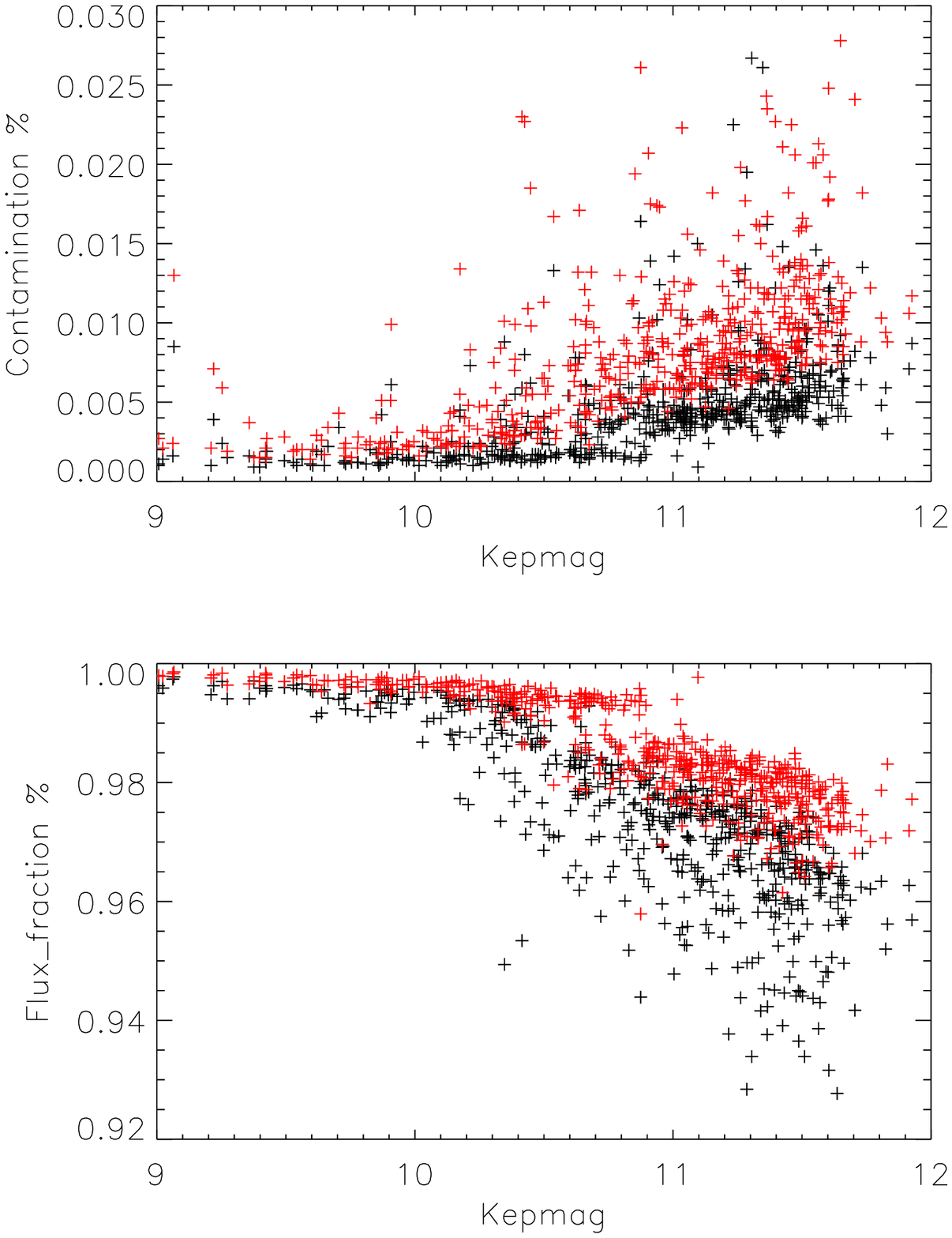}
\includegraphics[width=0.45\linewidth]{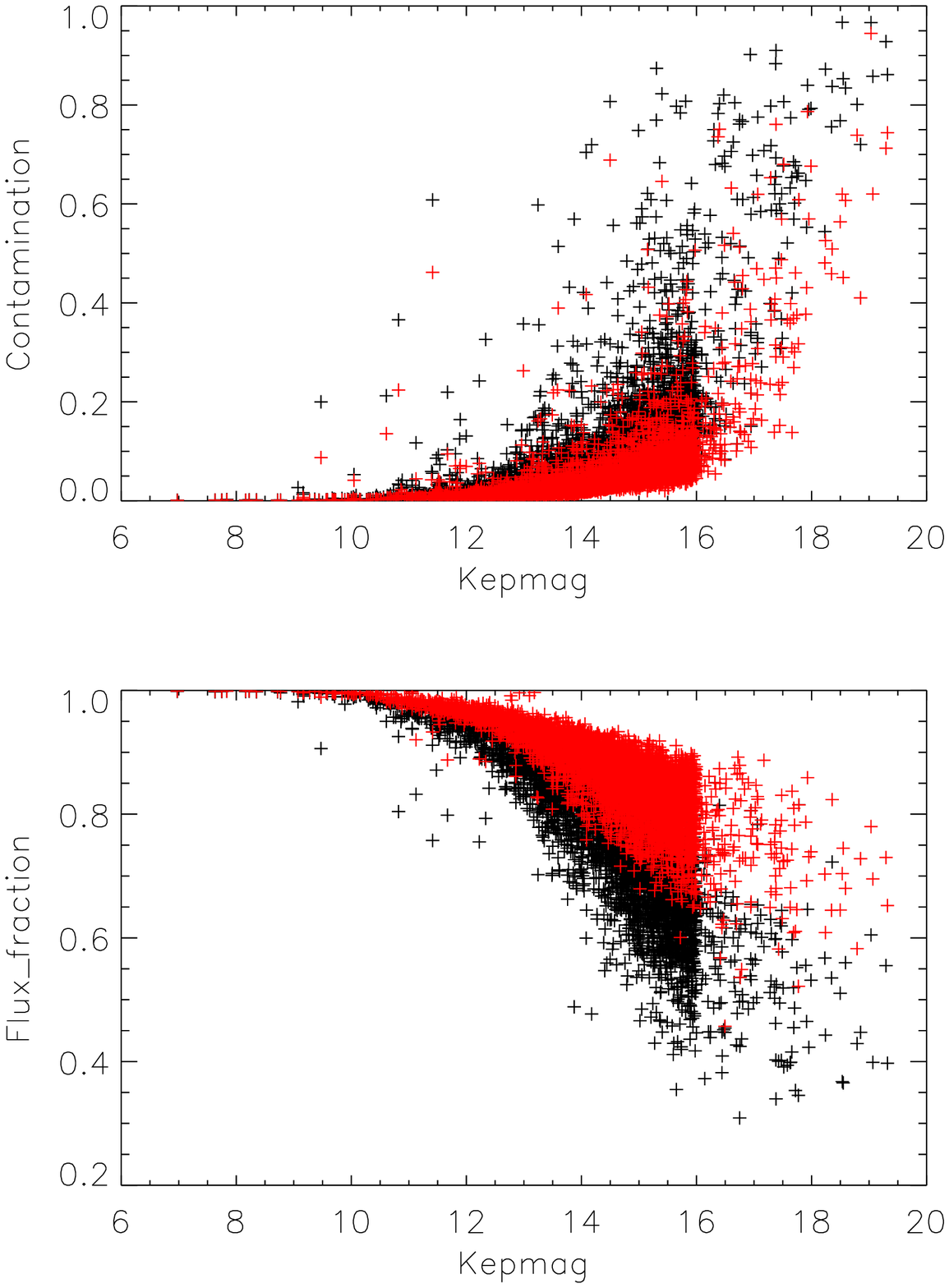}
\caption{\label{fig:fluxfrac}
Flux fraction and contamination for (left) asteroseismic calibration stars and (right) a larger sample of representative stars spanning a larger range of {\it Kepler} magnitudes. The two colors represent the maximum and minimum values observed over all quarters.  In general, the flux fraction (the fraction of a star's flux included in the photometric aperture) decreases with increasing magnitude while flux contamination from neighboring stars increases.  For the measurement of \fl, we exclude any quarters where the flux fraction is less than 0.9 and/or where the contamination is greater than 0.05.  Note the different y-axis scalings between the left and right columns: the photometry for the asteroseismic sample tends to be significantly cleaner than the larger, more representative sample.
}
\end{figure}

Figure~\ref{fig:fluxfrac} shows that the fraction of a star's flux that is included in the photometric aperture (flux fraction), and the fraction of the included flux that is due to neighboring stars spilling into the photometric aperture (contamination), are functions of the star's brightness but also can change from quarter to quarter.  Consequently, in the determination of the final \fl\ over all available quarters, we eliminate any quarters for which the flux fraction is less than 0.9 and/or for which the contamination is greater than 0.05. Note that these filters nearly always retain all available quarters for relatively bright stars ($K_p < 12$) but become increasingly important for fainter stars, where one or more quarters are usually excluded by these criteria.  We take as the final estimate of a star's $F_8$ the median or robust mean of that measured from the surviving quarters.

\subsection{Calibration data sets\label{sec:calib-ast}}

Following submission of \citet{bastien13}, a number of improved asteroseismic parameters for dwarf stars and subgiants were published, supplementing earlier works focusing on evolved stars.  We therefore update our calibration sample to include the best set of asteroseismic gravities currently available.  We draw our sample from the asteroseismic analyses of \citet{bruntt12}, \citet{thygesen12}, \citet{stello13}, \citet{huber13}, and \citet{chaplin14}, the latter two filling out the dwarfs in our sample.  This sample supercedes that of \citet{chaplin11a} used in \citet{bastien13}.  We note that the calibration stars used in \citet{bastien13} included some stars with poorly measured seismic parameters; the asteroseismic measurements of these objects are presumably negatively impacted by high levels of magnetic activity \citep{chaplin11b,huber11}.  We exclude these stars here. 

The new sample of 4140 stars considered here contains main-sequence, subgiant, and red giant stars, with \logg\ ranging from $\sim$0.5 to 4.56 dex; for the calibration itself, we restrict ourselves to stars with \logg$>$2.7 as stars more evolved than this deviate from the nominal relation (see Section~\ref{sec:db}).  We also exclude stars cooler than 4500 K, using the temperatures listed in the publications from which we draw our sample, and those with photometric ranges greater than 2.5 parts per thousand (ppt), as in \citet{bastien13}.  Note that we restrict ourselves here to stars with the smallest overall variability amplitudes in order to obtain the cleanest calibration, however as we show below we are able to apply the calibration to stars with somewhat larger variability amplitudes up to 10 ppt.  As described in, e.g., \citet{chaplin14}, surface gravities for many of these stars were determined via a grid-based approach coupling seismic observables ($\delta\nu$ and $\nu_{max}$) with independent measurements of \teff\ and [Fe/H].  In general, the uncertainty in the asteroseismic \logg\ values is $\sim$0.01 dex \citep[e.g.,][]{chaplin14}.  We note that this calibration sample contains stars with \teff\ values as hot as $\sim$7000 K, permitting us to extend applicability of \fl\ to stars hotter than the 6650 K limit of the previous relation reported in \citet{bastien13} (we discuss this further in Section~\ref{sec:teff_lim}).

\section{The Flicker Method}\label{sec:method}

\begin{figure}[ht]
\centering

\includegraphics[width=8.5cm]{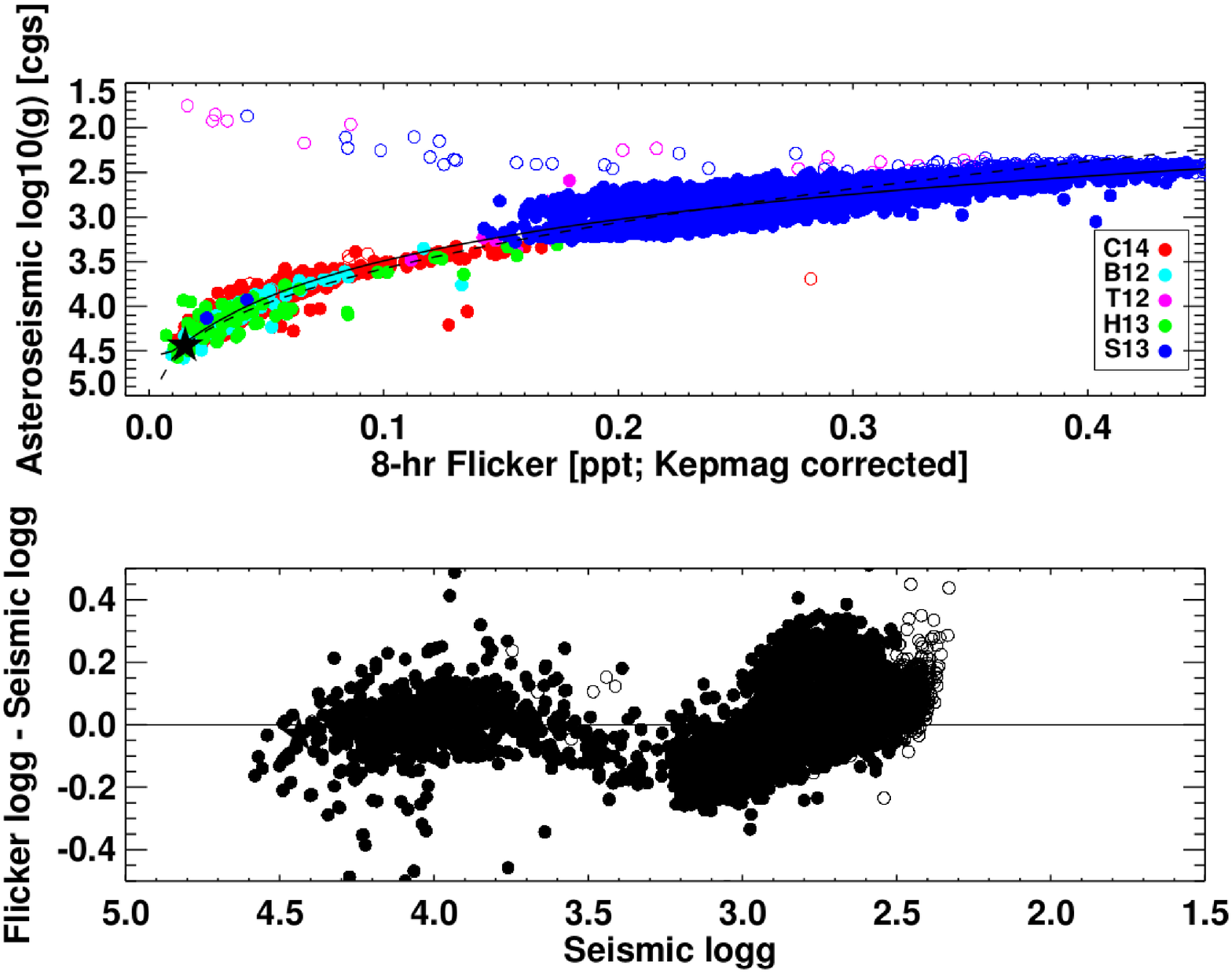}   
\newline     
\includegraphics[width=8.5cm]{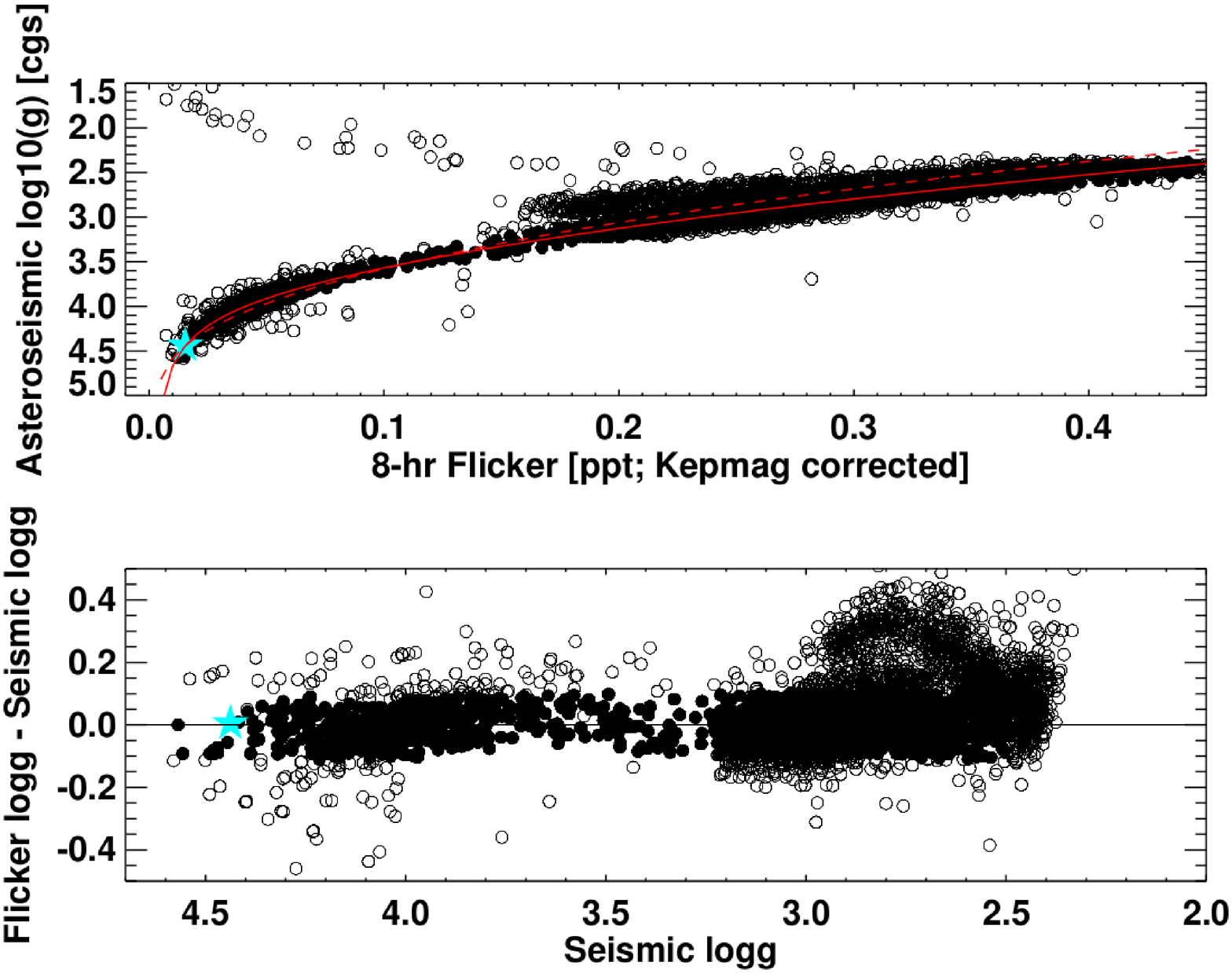}    

\caption{\label{fig:f8-logg-calib}
{\it Calibration of Flicker Against Asteroseismic \logg. }  {\it Top} 8-hr flicker vs. asteroseismic \logg\ for the calibration samples used in this work.  Red points are from \citet{chaplin14}, cyan points are from \citet{bruntt12}, and magenta points come from \citet{thygesen12}.  Data from \citet{huber13} are shown in green, and we plot the data from \citet{stello13} in blue; this last data set contains a mixture of red giants and red clump stars.  The star symbol to the lower left represents the Sun, through which we force our fit.  Solid points are those we used in the initial fit; open circles are ``double-backed'' stars that we exclude from the fit (see text).  The curves show our initial fits to the data; the residuals ({\it second from top}) show significant structure, caused by the double-backed stars pulling the fits.  The r.m.s. scatter about these residuals is 0.1 dex, and the median absolute deviation is 0.09 dex.  {\it Third from top: } To reduce the impact of the double-backed stars on our fit, we subsequently iteratively reject outliers lying 0.1 dex above and 0.2 dex below the polynomial fit, achieving convergence after 3 trials.  The dashed curve is our initial best fit, and the solid curve is the result of fitting after outlier rejection.  Open circles are points that were ultimately rejected from the fit while solid ones were retained.  This process not only removes the double-backed stars but also removes red clump stars, which show a slightly different \fl-\logg\ dependence (see Section~\ref{sec:db}).  The residuals, shown in the {\it bottom} panel, reveal significantly less scatter about the final fit for the stars used in the fit: the r.m.s. is 0.05 dex and the median absolute deviation is 0.04 dex, however we use the r.m.s. and median absolute deviation from our initial fit as our formal uncertainties for the \fl-based gravities.
}
\end{figure}

\subsection{Determining the best smoothing timescale}

To determine the best smoothing timescale, we computed flicker with smooths ranging from 1~hr to 18~hr (2pt to 36pt), following the general methodology described in \citet{basri11} and \citet{bastien13}.  For this, we focus on the \citet{chaplin14} sample, as it is the only one here that samples both dwarfs and more evolved stars well.  We find that the relationship between flicker and asteroseismic \logg\ holds well for all timescales considered, with the scatter about the relationship always less than $\sim$0.15~dex (see Table~\ref{tab:smooths}).  However, the number of dwarf outliers increases as the smoothing timescale increases, presumably largely due to the effects of magnetic activity (see \citet{bastien13}, but see also Section~\ref{sec:range_boosters} below).  For each smoothing timescale, we compared the flicker with asteroseismic \logg\ and fit a polynomial to the result as in Section~\ref{sec:calib}.  We determined the scatter about the fit by computing the r.m.s.\ and median absolute deviation, which we report in Table~\ref{tab:smooths} for each timescale examined.  The 4-hr smooth yields the smallest scatter, and thus the most robust \logg, for the widest range of \logg\ values.  However, the differences between the 4-hr and 8-hr smooths are small.  We also note that the sample used in \citet{bastien13} yielded the best performance with the 8-hr smooth.  We therefore adopt this as our smoothing length of choice for the sake of consistency with our previously published work.

\subsection{Doubling back\label{sec:db}}
Figure~\ref{fig:f8-logg-calib} reveals a subset of our asteroseismic calibration sample that deviates from the nominal \fl-\logg\ relation.  Instead, these stars follow a trend of decreasing \fl\ with decreasing \logg, effectively ``doubling-back'' in \fl-\logg\ space.  This is likely caused by the combination of stellar evolution and the chosen flicker smoothing window: as stars evolve, the timescales and amplitudes of both granulation and solar-like oscillations increase.  At some point, around \logg$\sim$2.7, the dominant granulation timescale begins to cross 8 hours into longer timescales excluded from the \fl\ metric, such that the contribution of granulation to the \fl\ signal decreases.  The timescales and amplitudes of the solar-like oscillations, on the other hand, remain comparatively small but increase to levels detectable with flicker, presumably becoming the dominant driver of flicker by \logg$\sim$2.

\fl, as currently defined, therefore suffers from a degeneracy, where highly evolved stars may masquerade as dwarfs and subgiants.  The highly evolved stars are readily identified through the Fourier spectra of their light curves, where they show clear solar-like oscillations indicating their evolved status.  This was used to place constraints on the evolutionary status of unclassified stars in the {\it Kepler} field \citep{huber14}.  Alternatively, one may use procedures like those outlined in \citet{basri11} (their Figure 6), which leverage the fact that very low gravity stars exhibit fluctuations on timescales that are longer than those probed by \fl\ and which are larger, slower, and more aperiodic than the starspot-driven variations of rapidly rotating dwarfs.  We applied this procedure to some of the double-backed stars and verified that it indeed adequately identified very low gravity stars.

Other methods can readily distinguish highly evolved stars from dwarfs and subgiants;  reduced proper motions are one example \citep{stassun14}.  Colors, such as those used in the KIC \citep{brown11} and for a significant number of stars in the updated {\it Kepler} stellar properties catalog \citep{huber14}, are also highly effective.  Hence, contamination from low \logg\ stars is relatively easy to eliminate, and we remove stars with \logg$<$2.5, as determined by \citet{huber14}, from the sample of stars with \fl-based gravities listed in Section~\ref{sec:f8tab}.

\begin{figure}[ht]
\includegraphics[width=8.5 cm]{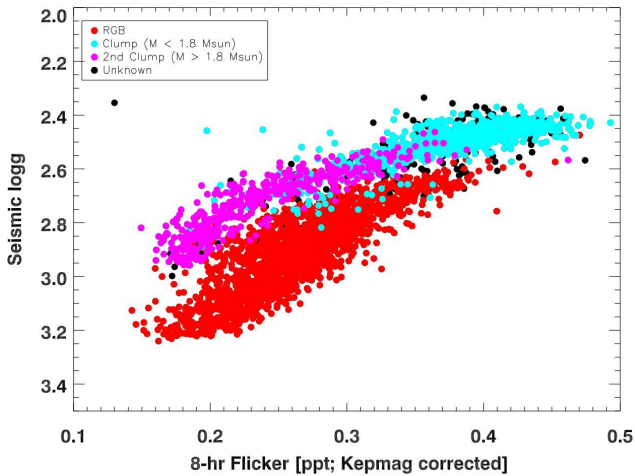}   
\caption{\label{fig:clump}
{\it Red clump stars and red giant branch (RGB) stars have different \logg\ values but similar \fl:}  Red clump stars (cyan and magenta points) cleanly separate from red giant stars (red points) in the \fl-asteroseismic \logg\ diagram, with clump stars having somewhat lower \logg.  Both populations, however, tend to have similar \fl\ values, suggesting similar contributions to the \fl\ signal from convective motions despite different evolutionary states.  Clump and secondary clump stars also tend to separate, with the lower mass clump stars tending to have lower \logg\ but larger \fl.  Black points represent stars whose evolutionary state is unknown \citep[cf.][]{stello13}; many of these lie on the red clump sequence of the \fl-\logg\ diagram, suggesting that \fl\ might be used as a constraint in addition to the asteroseismic analysis to help elucidate their nature.}
\end{figure}

We note one other interesting feature in Figure~\ref{fig:f8-logg-calib}: a clear population of stars with 2.5$\lesssim$\logg$\lesssim$3.0 and 0.15$\lesssim$\fl$\lesssim$0.3 that deviates from both the nominal and the double-backed relations.  These are red clump stars, as we show in Figure~\ref{fig:clump} where we zoom in on this region and color code the points by stellar mass.  Although there is some overlap between them, we see that we can even distinguish between clump and secondary clump stars.  That the clump stars overlap with the red giants in this region means that the \fl-\logg\ values of these stars can have systematic uncertainties of up to 0.2~dex.  However, the properties of such stars can readily be measured via asteroseismology even with long cadence light curves, perhaps reducing the value of a stand-alone \fl-\logg\ measurement.  We suggest, however, that combining the asteroseismic and \fl\ measurements can yield potentially interesting insights into the convective properties of red giant vs. red clump stars.

\subsection{Effect of activity, pulsations, and other range boosters\label{sec:range_boosters}}

We examined whether magnetic activity can influence or bias the \fl\ measurements and therefore the \logg\ estimates. We established in \citet{bastien13} that activity levels up to that of the active Sun do not affect the solar \fl\ amplitude. On the other hand, \citet{basri13} and other authors have shown that about a quarter of the {\it Kepler} stars are more active than the active Sun. It is true for all but the most rapid rotators that the timescales of the variability induced by magnetic activity are long enough that the 8 hour timescale used for \fl\ might be fairly insensitive to most effects.  Flares are a notable exception to this, however our sigma clipping procedures should remove most of these. 

In any case, it is worthwhile to investigate whether there is a correlation between \fl\ and an ``activity'' diagnostic. For the latter, we choose the ``range'' as defined by \citet{basri11}. This is a measure of the total differential photometric amplitude covered over a defined (usually long) timescale, typically a month or a quarter. In order to lower the sensitivity of this diagnostic to flares, transits, and other anomalies, we define the range as the amplitude between the 5\% and 95\% lowest and highest differential photometric points. Because of the types of stars and photometric behavior that exhibit high ranges, \citet{bastien13} restricted themselves to stars with ranges less than 3ppt. Indeed, there are rather few stars in the asteroseismic sample we consider that have larger ranges, so this sample is generally not very informative about possible activity effects. 

In \citet{bastien13} we did find some evidence for stars with Range $>$ 3ppt and Prot $<$ 3 days to be outliers. However, the \fl\ procedures used in that paper did not include clipping of flares, so it appears that problems with at least modest levels of activity have been ameliorated. Indeed, in \citet{bastien14b}, we showed that with clipping of flares and of transits, there is no noticeable effect on the agreement between \fl\ and asteroseismic \logg\ for Range as high as $\sim$10 ppt.

Given that the asteroseismic calibration sample include relatively few stars with large ranges, we considered another sample of stars: transiting exoplanet candidate host stars ({\it Kepler} Objects of Interest, KOIs) which have careful spectrscopic analyses of their gravities. Since these are mostly dwarfs, and some of them are relatively young, they provide is a better sampling of ranges. We are indebted to Erik Petigura and Geoff Marcy for sharing their compilation of gravities prior to publication (Petigura et al., in prep). Among these, there were 655 which had ranges less than 3ppt, 171 with ranges between 3-9ppt, and 81 with ranges greater than 9ppt (the maximum was 90ppt).  Using this sample, we were not able to discern any trends with range in the difference between \fl\ and spectroscopic \logg, either with apparent magnitude or effective temperature. There is a clear effect in the range itself with effective temperature (and equivalently gravity, since these are dwarfs); this is the previously reported tendancy of cool (higher gravity) dwarfs to show greater photometric variability \citep[cf.][]{basri13}.  We therefore find no convincing evidence that magnetic activity adds uncertainty to flicker gravities, at least not for dwarf and subgiant stars and not above the contribution from simple photon noise.

\subsection{Surface Gravity Calibration and Limitations\label{sec:calib}}

After preparing the light curves of the asteroseismic calibrators following the steps outlined in Section~\ref{sec:measf8}, we compute the \fl\ of each set of light curves.  This entails subtracting an 8-hr smoothed version of the light curve from itself and measuring the standard deviation of the residuals \citep[cf.][]{bastien13,basri11}, thereby providing a measure of the amplitude of the stellar variations occurring on timescales shorter than the smoothing timescale (here, 8 hours).  We compare these \fl\ values with the asteroseismically measued \logg\ values from the studies listed in secton~\ref{sec:calib-ast} and fit a polynomial to the data points with asteroseismic gravities greater than 2.5.  We force the fit through the solar gravity as it is the most well-constrained \citep[see][]{bastien13}.  As can be seen in Fig.~\ref{fig:f8-logg-calib} ({\it top}), low gravity stars ``double back'' in the \fl-\logg\ diagram (Section~\ref{sec:db}), and this doubling back begins at \logg$\sim$2.7.  Hence, \logg\ values obtained with the calibration reported herein may be unreliable for stars with gravities below this limit.  The resulting residuals have significant structure (Fig.~\ref{fig:f8-logg-calib}, {\it second from top}), caused primarily by the double-backed stars pulling the polynomial fit.  We therefore iteratively reject outliers that are 0.1 dex above and 0.2 dex below the polynomial fit. We achieve convergence in the fit after three iterations; Fig.~\ref{fig:f8-logg-calib} ({\it third from top}) shows the resulting fit, and the {\it bottom} panel shows the significantly reduced scatter about the fit for the remaining stars used in the fit.

The final calibrated fit, updated from and superseding that of \citet{bastien13}, is:
\begin{equation}
\log g = 1.3724221 - 3.5002686x - 1.6838185x^{2} - 0.37909094x^{3}
\end{equation}

\noindent
where $x = log_{10}(F_{8}) $ and \fl\ is in units of ppt.  The r.m.s.\ of the \logg\ residuals about the initial fit is 0.1 dex, and the median absolute deviation is 0.09 dex.  These represent conservative uncertainties that we assign to \fl-\logg\ values (but see Section~\ref{sec:faint_lim}).
We note that red clump stars (discussed in section~\ref{sec:db}, above) are removed from our calibration, but in real observations, where the \logg\ is not known a priori, the presence of red clump stars results in a systematic error in the \fl-\logg\ of red giant stars. 
The sense of the systematic error is that a red giant with an \fl-inferred \logg\ in the range $\sim$3.05$\pm$0.15 could potentially have a true \logg\ that is up to $\sim$0.3 dex lower.
Because the range of \logg\ affected by this potential systematic error is so narrow and furthermore does not change the inferred luminosity class of a star, we do not attempt to include this systematic error in our reported \fl\ \logg\ values, which as discussed show an overall tight relation with $\sim$0.1 dex error.

\subsubsection{SAP versus PDC-MAP\label{sec:sap_vs_pdc}}

The {\it Kepler} mission produced two different pipeline products, either of which may in principle be used to measure \fl.  The calibration reduction (SAP) has the virtue that it has not been subjected to filtering on various timescales by the PDC-MAP procedures, though we do not necessarily expect signifcant filtering from these procedures on the timescales of interest to the flicker calculation. On the other hand, the SAP light curves have the disadvantages of including various instrumental effects (some of which are on relevant timescales) that PDC-MAP has removed, as well as jumps in the contiuum level that are also removed by the downstream pipeline (although these should not have much influence given the way we calculate \fl). To assess which of the two producs, SAP or PDC, is preferred for the \fl\ measurement, we tested the two methods on our asteroseismic calibration sample.

We found that both of the light curve products produced \fl-\logg\ values with similar scatter, but that those from SAP curves had a general offset in the inferred gravities of about +0.08 dex (i.e. SAP-inferred gravities were generally higher by that amount). This is a bit surprising, since one might have expected that the extra instrumental signal left in SAP light cuves (heater events and the like) would produce additional short-term variability in the light curves, which would yield higher flicker amplitudes and therefore lower gravities. Less surprisingly, we found that the offsets were smaller for stars with lower gravity or brighter magnitudes (which would tend to have a higher signal-to-noise ratio for a flicker measurement). This offset results if one uses a single calibration curve for both analyses; for instance, if the PDC-MAP light curve are use to first calibrate the flicker relationship, and that calibration is then applied to both SAP and PDC-MAP lightcurves. However, one can construct different calibration curves for each of these reductions against the same seismic sample, thus removing the offset. As users will likely prefer to use the cleaner PDC-MAP light curves, we use these in this work.

\subsubsection{Faintness limits vs.\ gravity\label{sec:faint_lim}}

A fundamental limitation of the flicker method is that the observed \fl\ signal becomes increasingly dominated by shot noise as one considers fainter stars. While the methodology described here allows the detection of gravity-sensitive granulation signals for stars that are considerably fainter than those that can typically be studied asteroseismically, the shot noise does ultimately swamp the granular signal as well. Moreover, because the \fl\ amplitude of the granular signal is smaller for higher \logg\ stars, the shot noise becomes more quickly dominant at higher \logg.

To quantify this, we use our empirical polynomial relation for the shot noise contribution as a function of {\it Kepler} magnitude ($K_p$; Eq.~\ref{eq:f8corr}), together with our asteroseismically calibrated \fl-\logg\ relation, to estimate the maximum \logg\ at a given $K_p$ that produces a granulation \fl\ signal that is above a certain fraction of the shot noise. Table~\ref{tab:logglimit} summarizes the \logg\ values representing 100\%, 20\%, and 10\% of the shot noise as a function of $K_p$.

As can be seen, requiring the intrinsic flicker to be at least as large as the shot noise would imply that flicker is only sensitive down to $K_p \sim 9.0$ for a solar type \logg\ of $\sim$4.4. It is clear, however, that the method can be applied reliably much fainter than this; the seismic calibration sample includes stars with solar \logg\ as faint as 11.7, and those are recovered with an accuracy of $\sim$0.1 dex. This suggests that we can reliably extract flicker signals that are as low as 20\% of the shot noise (the \logg$_{20\%}$ column shows that \logg\ of 4.4 produces flicker that is 20\% of noise at $K_p$~$\sim$12.5). It is possible that one can reliably extract flicker signals that are as low as $\sim$10\% of the shot noise, meaning solar type \logg\ for stars as faint as $\sim$13.5. Unfortunately, the seismic calibration sample does not include such high \logg\ stars fainter than 12.5, so we cannot test this directly.

Therefore, as a conservative estimate, we assume that intrinsic granulation \fl\ signals that are less than 20\% of the shot noise are not reliably recoverable. Since the true \logg\ of a main sequence star can be as high as $\sim$4.5, we therefore adopt an uncertainty on the \fl-based \logg\ that is the difference between the \logg$_{20\%}$ value in Table~\ref{tab:logglimit} and 4.5. For example, at $K_p = 12.5$ the uncertainty on \logg\ is $\sim$0.1 dex, whereas at $K_p = 13.5$ the \logg\ uncertainty is as high as $\sim$0.3 dex. Note that the uncertainty is asymmetric: flicker provides a lower limit on the gravity but not a meaningful upper limit. For such faint stars, additional \logg\ estimates from, e.g., spectroscopy (given a sufficient signal-to-noise and well-calibrated method) can provide potentially tighter constraints on \logg. Meanwhile, if the flicker methodology can be reliably extended to extract reliable signals down to 10\% of the shot noise, it should be possible to achieve $\sim$0.1 dex accuracy on \logg\ for main sequence stars as faint as $K_p \sim 14$.

\subsubsection{Temperature limits\label{sec:teff_lim}}

In this work, we benchmark \fl\ against the most reliable stellar parameters available for stars in the {\it Kepler} field, and hence the present calibration is restricted by the limits of this sample.  This sample includes stars significantly hotter than 6650 K, where a substantial outer convective zone is generally not expected.  Still, the \fl\ calibration holds even for these stars.  Meanwhile, \citet{kallinger10} find tantalizing evidence for granulation in stars with spectral types as early as early A, meaning that a \fl-like relation could perhaps be extracted for even these stars, assuming they do not pulsate.  We note that the actual, current limits of \fl\ on the hot end are unclear: many of the effective temperatures of the stars with temperatures $>$7000K in the asteroseismic sample were derived from broad-band photometry, but some initial spectroscopic re-analyses of these stars suggest effective temperatures closer to $\sim$6800K (D. Huber, private communication).  Here, we set our limits based on the published effective temperatures.  Also, although we find that we recover \logg\ to within 0.1 dex of the asteroseismic gravity in these stars, we caution below that the true uncertainty in the early F stars may be larger.  Nonetheless, if granulation is the primary driver of flicker \citep[cf.][]{cranmer14,kallinger14}, then this methodology may in principle be applied to any star exhibiting surface convection, and so any star with an outer convective zone.  As such, possible applications of the flicker method may include Cepheids \citep{neilson14}, M dwarfs, and perhaps even white dwarfs.

As an initial test of this hypothesis, we calculated the \fl\ of the the bright K-dwarf recently analyzed by \citet{campante15}.  We note that the asteroseismic calibration samples we use here and in \citet{bastien13} do not contain any K dwarfs, primarily due to the difficulties in measuring the low amplitude asteroseismic signals of such stars until now.  The asteroseismically determined \logg\ of the star studied by \citet{campante15} is 4.56$\pm$0.01, whereas the $F_8$ based \logg\ is 4.52$\pm$0.1, in very good agreement.  Our ability to recover this star's very low \fl\ signal is primarily due to its brightness ($K_p = 8.7$). Nonetheless, this test case indicates that from a physical standpoint the granular \fl\ signal can be used to accurately measure \logg\ for early-K dwarfs, and that the \fl-\logg\ calibration derived here may be applied to such stars.

\subsubsection{Consistency of flicker gravities across quarters}

\begin{figure}[!ht]
\includegraphics[width=8.5cm]{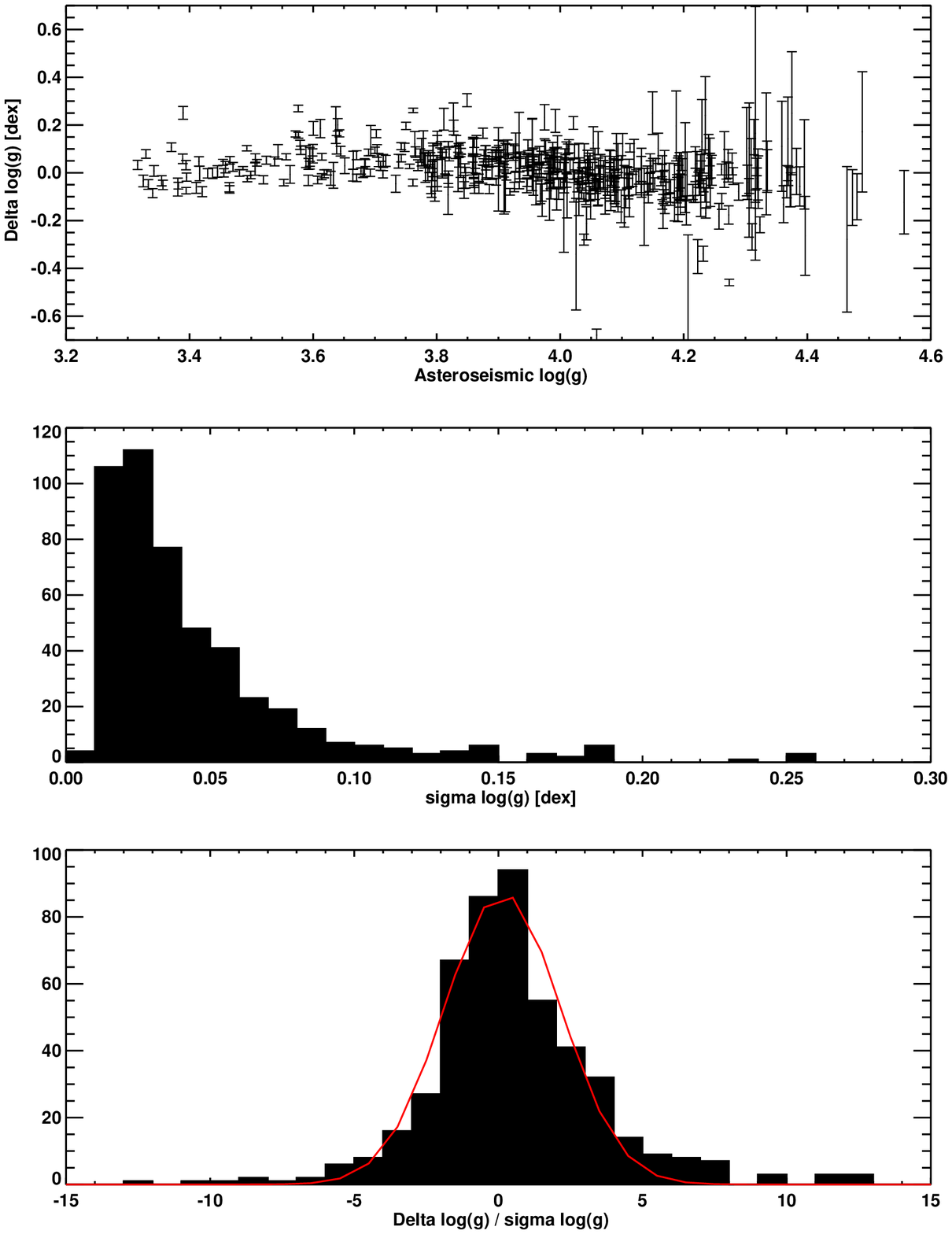}   
\caption{{\it Assessment of quarter-to-quarter variations in \fl-based \logg\ using the asteroseismic calibration sample from \citet{chaplin14}}. {\it Top:} Residuals of \fl-based \logg\ relative to asteroseismic \logg, with error bars representing the r.m.s.\ standard deviation of the \fl-based \logg\ across all available quarters. These error bars are in general smaller than the absolute scatter in the residuals relative to the asteroseismic benchmark. {\it Middle:} Distribution of the quarter-to-quarter r.m.s.\ scatter in the \fl-based \logg. The typical quarter-to-quarter variation in \fl-based \logg\ is $\sim$0.02--0.05 dex. {\it Bottom:} Distribution of the ratio of \logg\ residuals to the quarter-by-quarter r.m.s. The distribution is reasonably approximated by a Guassian with $\sigma = 2.0$, indicating that the true \fl-based \logg\ errors are a factor of 2 larger than the relative quarter-by-quarter errors for a given star.}
\label{fig:sigvq}
\end{figure}

We showed above that our calibration of \fl\ against a sample of asteroseismically measured \logg\ has a typical r.m.s.\ scatter of $\approx$0.1 dex for bright stars with $K_p < 12.5$. Interestingly, we find that the relative quarter-to-quarter variations in \fl-based \logg\ are considerably smaller than the 0.1 dex absolute deviations. We show this in Figure~\ref{fig:sigvq} for the subsample of asteroseismic calibrators from \citet{chaplin14}, which span nearly the full range of \logg\ for which \fl\ is applicable, and all of which have $K_p < 12.5$. The top panel displays the residuals between \fl-based and asteroseismic \logg, where the error bars represent the standard deviation of each star's quarter-by-quarter \fl-based \logg. For many stars the error bars are smaller than the spread of the residuals, and this holds true over the full range of \logg. The middle panel displays the distribution of the quarter-by-quarter standard deviations in \fl-based \logg, where we see that indeed the \fl-based \logg\ is very stable from quarter to quarter, with a typical r.m.s.\ of 0.02--0.05 dex. The bottom panel represents the distribution of the ratio of stars' residuals relative to their quarter-by-quarter r.m.s. This distribution is reasonably well approximated by a Gaussian with $\sigma = 2.0$, indicating that for the typical star in the calibration sample, the quarter-to-quarter variations in \fl-based \logg\ are a factor of $\sim$2 more stable than is the absolute deviation of the stars' \fl-based \logg\ relative to the true asteroseismic \logg.

 One possible reason for this is that the absolute deviations of $\sim$0.1 dex in \fl-based \logg\ are at least partly driven by other, as-yet unmodeled parameters that act to inflate the \fl-based \logg\ errors, as the surface gravity does not, by itself, set the granulation amplitude.  Indeed, the amplitudes may be affected by the stellar mass \citep{kallinger14}, magnetic activity \citep{huber11}, or metallicity.  As such, if these other parameters can be identified and included in the calibration, then it should be possible to reduce the absolute \fl-based \logg\ errors to $\sim$0.03 dex (Fig.~\ref{fig:sigvq}, {\it middle}).

\section{Results}\label{sec:results}

\subsection{Flicker Gravities of {\it Kepler} Stars\label{sec:f8tab}}

We provide in Table~\ref{tab:f8} \fl-based gravities for the 27\,628 {\it Kepler} stars that satisfy the criteria listed in Section~\ref{sec:method}.  Of these, 6062 stars are brighter than $K_p =$12, with 470 brighter than $K_p =$10.  The sample contains $\sim$8328~K stars, 6605~G stars, and 12\,695~F stars (of which 2365 are early F stars).  As discussed below, we find that a significant fraction of the sample consists of subgiants.  We include in this table stars with asteroseismically measured gravities to encourage further comparisons between \fl\ and asteroseismology.

\subsection{Comparison between Flicker Gravities and Expectations from Stellar Population Synthesis Models}

The \logg\ values for the {\it Kepler} stars inferred from the granulation flicker (Table~\ref{tab:f8}) span the range 2.5 $<$ \logg\ $\lesssim$ 4.6 as expected for stars representing main sequence, subgiant, and ascending red giant branch evolutionary stages. The upper bound on \logg\ presumably simply reflects the \logg\ corresponding to main sequence stars at the cool end of the \fl\ calibration (\teff\ $\approx$ 4500 K) while the lower bound on \logg\ is an artificial cutoff imposed by the limits of the \fl\ calibration (see Sections~\ref{sec:db} and~\ref{sec:calib}). 

In particular we note that the distribution of \fl\ \logg\ values is not sharply concentrated around main sequence values (\logg\ $\gtrsim$ 4.1) but rather shows a broad distribution in the range 3.5 $<$ \logg\ $<$ 4.5, and in particular includes a significant population with 3.5 $<$ \logg\ $<$ 4.1: subgiants. Indeed, among the stars with inferred \logg\ $>$ 3.5 (i.e., excluding more evolved red giants), subgiants apparently constitute $\sim$60\% of the {\it Kepler} sample studied here. This may be contrary to intuition considering that (a) subgiants are intrinsically rare compared to main sequence stars in the overall underlying Galactic disk population and (b) the {\it Kepler} target prioritization specifically attempted to prioritize small main sequence dwarfs \citep{batalha10}.

\begin{figure}[ht]
\includegraphics[width=8.5cm]{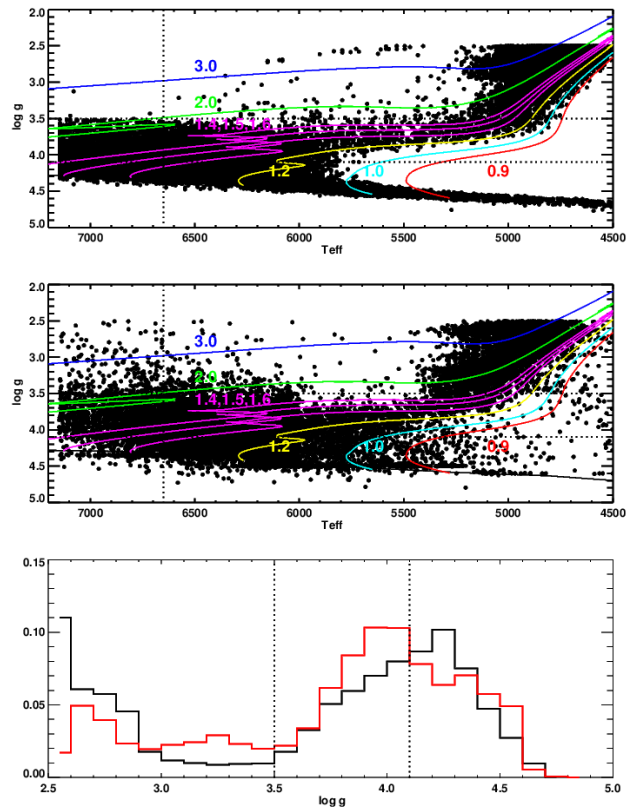}   
\caption{{\it \fl-based \logg\ for} Kepler {\it stars in the H-R diagram.} {\it Top:} Simulated population from the TRILEGAL model for the {\it Kepler} field of view and down to $K_p < 13.5$, with evolved red giants (\logg\ $<$ 2.5) removed and the \teff\ range restricted to 7150 K $>$ \teff\ $>$ 4500 K to mimic the limits of the \fl\ calibration. We also show evolutionary tracks for stars of various masses, indicated in units of \msun. {\it Middle:} Same, except showing the actual \fl\ based \logg\ for the {\it Kepler} sample. {\it Bottom:} Comparison of the distributions of \logg\ from \fl\ (red) and TRILEGAL simulation (black). The vertical dotted lines indicate the range of 3.5 $<$ \logg\ $<$ 4.1, corresponding to subgiants.
\label{fig:trilegal}
}
\end{figure}

It is useful therefore to consider how the \fl\ based \logg\ values compare to what might be expected from standard Galactic stellar population synthesis models.  Except for the deliberate exclusion of highly evolved red giants, the {\it Kepler} target sample is expected to be representative of the field for $K_p \lesssim 14$ \citep{batalha10}; the target sample we consider here satisfies this as we have considered only stars with $K_p < 13.5$.  Figure \ref{fig:trilegal} (top panel) shows the results of a simulated {\it Kepler} sample in the H-R diagram plane produced with the TRILEGAL model \citep{girardi05}. We used the default TRILEGAL parameters, and simulated the {\it Kepler} field of view by generating 21 pointings each of (5 deg)$^2$ corresponding approximately to the positions and sizes of the 21 Kepler CCD pairs. We also restricted the simulated stellar sample to $K_p < 13.5$, \logg\ $>$ 2.5, and 7150 K $>$ \teff\ $>$ 4500 K, so as to mimic the {\it Kepler} sample studied here according to the limits on the \fl\ calibration.

The simulated sample shows several features of interest. First, for \teff\ $\lesssim$ 4800 K, stars clearly bifurcate in \logg\ such that they are either unevolved cool dwarfs or evolved red giants. There is an expected ``no man's land" in \logg\ between these two groups where no stars are expected, essentially because there are not expected to be subgiants corresponding to evolved stars less massive than $\sim$0.9 \msun, since such stars have main sequence lifetimes longer than the age of the Galaxy. Second, there are virtually no stars with \logg\ $<$ 3.5 at \teff\ $\gtrsim$ 6650 K. Such stars would correspond to intermediate mass (2--3 \msun) subgiants rapidly crossing the Hertzsprung gap. Such a population does exist at \teff\ $\lesssim$ 6500 K, as the stars approach the base of the red giant branch. Third, stars with \teff\ intermediate to these cool and hot extremes, representing the majority of the sample considered here, fill a broad but well defined ``swath" with 3.5 $<$ \logg\ $<$ 4.5, representing main sequence stars and subgiants with apparent masses of 1--2 \msun.

The H-R diagram for the actual {\it Kepler} stars using the \fl\ based \logg\ values is shown for comparison in Figure \ref{fig:trilegal} (middle panel). For visual simplicity we do not show the \logg\ error bars, however it is important to bear in mind the asymmetric nature of the \fl\ \logg\ errors (see Section~\ref{sec:faint_lim} and Table~\ref{tab:f8}); we revisit the impact of the \fl\ \logg\ errors below. Broadly and qualitatively speaking, there is good agreement between the actual and simulated H-R diagrams. There are two main differences that appear. First, at the hot end (\teff\ $\gtrsim$ 6650 K) there is clearly a larger than expected population of apparently intermediate mass stars crossing the Hertzsprung gap, and this cannot be explained by the \fl\ \logg\ errors. We suspect that there may be other sources of variability contaminating the granulation flicker signal for a subset of stars with \teff\ $>$ 6650 K. Thus, even though the \fl\ \logg\ performs very well for the asteroseismic calibration sample which includes a number of stars as hot as 7150 K (see Section~\ref{sec:calib-ast}), we recommend that the \fl\ \logg\ values for such hot stars be regarded with caution. Second, there is an apparent population of cool stars with \teff\ $\lesssim$ 4800 K in the ``no man's land" discussed above, between the main sequence and red giant branch. In this case, the discrepancy is readily understood in terms of the asymmetric nature of the \fl\ \logg\ errors, which for these cool and mostly faint stars can be large and therefore make the \fl\ \logg\ for these stars consistent with the main sequence to within 1--2$\sigma$ in most cases (see also below).

As in the simulated H-R diagram, the actual sample shows a broad but well defined population of apparent main sequence and subgiant stars filling the region 3.5 $<$ \logg\ $<$ 4.5, again indicating a large population of mildly evolved subgiants in the sample, with apparent masses of 1--2 \msun. Figure \ref{fig:trilegal} shows the distributions of \logg\ for the simulated TRILEGAL sample (black) and the \fl\ based \logg\ for the actual {\it Kepler} sample (red).  Overall the broad agreement between the two distributions is encouraging.  We do not concern ourselves here with the differences at \logg\ $<$ 3.5, since as discussed above the {\it Kepler} sample is by design not representative of the field at these low gravities. The mild excess of stars in the \fl\ distribution at 3.0 $<$ \logg\ $<$ 3.5 is a further sign of potential problems with the \fl\ \logg\ for some hot stars with \teff\ $>$ 6650 K, also mentioned above.

However, we highlight here two features of the \logg\ distributions at \logg\ $>$ 3.5, where we expect the {\it Kepler} sample to be representative of the field. First, there appears to be a slight excess of very high-gravity dwarfs with \logg\ $\gtrsim$ 4.5. This is at least partly due to the fact that the \fl\ \logg's are not in any forced to match the expected main sequence, and thus in some cases scatter below it, i.e., to higher \logg. Second, there is an apparent offset in the peaks of the two distributions around \logg\ $\approx$ 4.1, such that there is an apparent excess of subgiants with 3.7 $<$ \logg\ $<$ 4.1 in the \fl\ \logg\ distribution relative to the simulated distribution. Indeed, among all stars with \logg\ $>$ 3.5, the subgiant fraction in the \fl\ distribution is 60.6\% as compared to 47.1\% from the simulated distribution. This is a manifestation of the asymmetric errors on the \fl\ \logg\ values, which are more likely to scatter the \logg\ to lower values than to higher values. In fact, adjusting the \fl\ \logg's by 1$\sigma$ results in a subgiant fraction of 48.7\%, in much closer agreement with the fraction from the simulated distribution.  In any event, it is clear that subgiants constitute a large proportion of the {\it Kepler} sample, at least for the bright sample ($K_p < 13.5$) considered here.

\section{Discussion}\label{sec:discussion}

\subsection{Theory of Why Flicker Traces Surface Gravity\label{sec:theory}}

As stated in section~\ref{sigclip}, \fl\ measures the r.m.s. of the stellar intensity variations with timescales shorter than 8-hr.  Earlier works, such as \citet{ludwig06}, showed that the total r.m.s. of the brightness variations, presumably driven by convective cells on the stellar surface, should scale as the square root of the number of granules \citep[see also][]{kallinger14}.  The seminal work of \citet{schwarzschild75}, focused on red giants, posited that the number of convective cells on the stellar surface, and hence the typical size of granules, is proportional to the pressure scale height, which in turn varies inversely with the stellar surface gravity.  More recent modeling efforts \citep[e.g.,][]{trampedach13} also find that the granule size strongly depends on the stellar evolutionary state, varying inversely with \logg.  Hence, the theoretical underpinnings of the \fl-\logg\ relation are fairly well-established.  In practice, though, it was unfeasible to extract the \fl\ signal from time-series photometry until the advent of missions like CoRoT and {\it Kepler}; attempts to model and observe the effects of granulation on spectral lines date at least as far back as \citet{dravins87}.  Additionally, as we describe above, the measurement of \fl\ can be complicated by effects like activity, as well as exoplanet transits (which become important at {\it Kepler}'s level of photometric precision) and other kinds of stellar and instrumental variability.  The smoothing on a particular timescale therefore serves as a filter to remove long-timescale variations unrelated to the granulation signal, and the additional clippings we perform help to mitigate the effects of shorter timescale variability distinct from granulation.

While the relationship between \fl\ and \logg\ was demonstrated in \citet{bastien13}, it was \citet{cranmer14} who first convincingly showed that granulation is the primary driver of the \fl\ signal by comparing the measured \fl\ signal with expectations for intensity fluctuations due to granulation from the models of \citet{samadi13a,samadi13b}.  As part of this study, \citet{cranmer14} also proposed a resolution to the discrepancy between the observed granulation amplitude in F stars and that predicted by standard granulation models.  This discrepancy is clearly observed using \fl, where the hotter stars are expected to exhibit faster convective motions, and so larger granulation amplitudes, but the observed amplitudes are significantly and systematically smaller than theory predicts.  However, by introducing in the models a term to suppress the granulation velocity field that depends on the stellar effective temperature (and so nominally on the depth of the outer convective zone), the authors were able to bring theory and data into agreement and thus highlight the possible importance of accounting for magnetic activity in granulation models of F stars.

Other approaches have also helped to confirm the relationship between \fl\ and granulation.  In particular, \citet{kallinger14} show that \fl\ almost perfectly matches the granulation amplitude measured in the asteroseismic power spectrum for stars with \logg$\lesssim$3.7; for higher gravities, they find that \fl\ begins to underestimate the granulation amplitude.  Nonetheless, this work serves as an observation-based demonstration that \fl\ traces granulation in addition to the theoretical treatments listed above.

\subsection{Some Applications of Flicker}

While \fl\ has been useful for stellar parameter estimation, its applications extend beyond the simple measurement \logg\ and stellar density \citep{kipping14}.  As discussed above in Section~\ref{sec:theory}, it may be used to help constrain models of convection, and comparisons with asteroseismology may yield useful empirical insights into the convective properties of stars.  Combining \fl\ with other measures of light curve variability \citep{basri11,basri13}, may permit novel probes into stellar evolution \citep{bastien13} and the interplay between magnetic activity and convection \citep{cranmer14}.

\fl\ has potential applications in exoplanet science as well.  Of particular interest is the ability to distinguish between subgiants and dwarfs, an issue of increasing concern in photometric surveys monitoring hundreds of thousands of stars such as {\it Kepler} and eveutually TESS \citep[see, e.g.,][]{brown11}.  Analysis of the discovery light curves obtained by these missions can also complement radial velocity follow-up campaigns by providing estimates of the radial velocity ``jitter'' for both magnetically active and inactive stars \citep{aigrain12,bastien14a} before deployment of ground-based telescopic resources.  Finally, \fl\ may aid in the ensemble characterization of large numbers of exoplanets not only through the measurement of \logg\ but also through methods such as astrodensity profiling \citep{kipping14}.

\subsection{Potential uses for upcoming missions}

While we have focused our efforts here on the {\it Kepler} field, flicker may in principle be used in other current and upcoming surveys.  Wide-field photometric surveys, such as the Palomar Transient Factory \citep[PTF;][]{law09}, Pan-STARRS \citep{jewitt03}, and the Large Synoptic Survey Telescope \citep[LSST;][]{becker07} are becoming a larger part of astronomical science and would initially seem to be potential sources of flicker data.  However, flicker relies on obtaining cadences ranging from a few hours to almost a day and also achieving photometric precision of 10 to 450 parts per million r.m.s.  In general, such capabilities are only possible for space-based telescopes.

Fortunately, several such missions are underway or being planned.  The revived {\it Kepler} satellite is now operating the K2 mission \citep{howell14}, observing fields along the ecliptic for $\sim80$ day durations over the next two years.  While the photometric precision of K2 is significantly poorer than {\it Kepler}, it should in principle be able to tease out the \fl\ signal 
\citep[e.g.,][]{vanderburg14}.  Over the projected lifetime of K2, the mission should be able to acquire flicker measurements of at least as many stars as did the prime {\it Kepler} mission.

Two upcoming missions have even greater promise for exploiting flicker.  NASA's Transiting Exoplanet Survey Satellite \citep[TESS;][]{ricker15} and the European Space Agency PLATO mission \citet{catala08} are being designed to detect transiting exoplanets with space-based photometric telescopes, but to do so over most or all of the sky.  TESS is planned to launch in 2017, to survey the entire sky, and to observe stars for 30 to 180 days at a time.  It will be mostly limited to bright stars ($I<12$).  PLATO will observe a large fraction of the sky with a longer time baseline than TESS, and with the ability to observe fainter stars than TESS.

Both TESS and PLATO will photometrically measure stars at relatively high cadence (2 minutes for the primary TESS targets and 25 seconds for PLATO).  In addition, TESS will acquire observations at a 30-minute cadence for everything in the sky.  Current plans for the mission show that roughly 500,000 dwarf stars across the sky should be observed with high enough precision for flicker measurements.  In general, each mission should be able to measure flicker for most of the bright stars they will probe for exoplanets, yielding many independent checks on the host star radii (via \logg) and thus the planet radii.

\section{Conclusion}\label{sec:conclusion}

In this paper, we provide an update to, and a more in-depth explanation of, the granulation ``flicker'' (\fl) relation published in \citet{bastien13}.  We now calibrate \fl\ against a larger and more robust set of asteroseismically measured \logg\ values.  We describe how we mitigate the adverse effects of astrophysical signals unrelated to granulation on the measurement of \fl, primarily via sigma clipping and removal of extrasolar planet transits.  We also perform a more robust removal of shot noise to the \fl\ signal than in \citet{bastien13}.  The shot noise removal, however, is imperfect, so we supply updated error bars for the \fl\ gravities.  These error bars are now asymmetric and are primarily meant to capture the increasing uncertainty in \fl\ with increasing magnitude; this generally means that the error bar is larger in the direction of higher \logg, i.e., toward the main sequence.  We explore the limitations of the technique, including faintness, temperature, activity and \logg\ limits.  In general, we find that \fl, as currently defined, may be measured in stars with 4500$<$\teff$<$7150, 2.5$<$\logg$<$4.5, $K_p>$13.5, and photometric ranges $<$10ppt.  At present, we exclude stars with known transiting extrasolar planets and known eclipsing binary stars.
We also caution that the application of the flicker methodology should be done with care for stars hotter than $\sim$6650 K.

We also uncover some intriguing areas for future exploration.  \fl\ as currently defined may only be reliably applied to stars with \logg$>$2.5, but it may be re-cast so that an \fl-like signal may be measured in stars with \logg\ as low as 1.5, or perhaps even lower.  While such stars are readily amenable to asteroseismic investigation, the additional measure of an \fl-like quantity may permit more detailed investigations into the convective properties of such evolved stars.  We find that, while red giants and red clump stars tend to have different \logg\ values, they have similar \fl\ values, for reasons that, to our knowledge, are not quite clear.  One possibility is that red clump and red giant stars with similar \logg\ have different \teff, which in turn affects the oscillation and granulation amplitudes and timescales (since their properties also depend on \teff).  We also extend the applicability of \fl\ to stars significantly hotter than 6500K, a result that can help to shed light on the surface properties of stars near the Kraft break \citep{kraft67,schatzman63}.  Additionally, we see that we might be able to improve the precision of the \fl-based \logg\ measurements to $\sim$0.03 dex, since in general the quarter-to-quarter \logg\ estimates are internally consistent to this level; we suggest that one or more secondary parameters are currently unaccounted for in our calibration, and if they can be identified and modeled, the true potential precision of the method could be recovered. Even so, a precision of $\sim$0.1 dex can already be achieved for most stars brighter than $K_p \sim 12.5$.

Ultimately, we provide a table of \fl-based \logg\ values and their errors for 27\,628 {\it Kepler} stars within the current parameters of applicability as a resource to the community for both exoplanet and stellar astrophysics investigations.

\acknowledgments
We thank Erik Petigura and Geoff Marcy for sharing their data with us in advance of publication, as well as Daniel Huber and Tom Barclay for helpful comments and discussions. We also thank the referee for helpful comments that improved the quality of the paper.  Support for this work was provided in part by NASA through Hubble Fellowship grant \#HST-HF2-51335 awarded by the Space Telescope Science Institute, which is operated by the Association of Universities for Research in Astronomy, Inc., for NASA, under contract NAS5-26555.

\begin{deluxetable}{lccc}
\tablewidth{0pt}
\tabletypesize{\scriptsize}
\tablecolumns{4}
\tablecaption{\label{tab:smooths}
Comparison between Different Smoothing Lengths}
\tablehead{ \colhead{Smoothing Length} &
            \colhead{Stdev of All} &
            \colhead{Robust Stdev} &
            \colhead{MAD}
          }
\startdata
2pt (1hr)       &  0.115  &  0.107  &  0.071  \\
4pt (2hr)       &  0.099  &  0.091  &  0.061  \\
8pt (4hr)       &  0.096  &  0.087  &  0.058  \\
16pt (8hr)      &  0.105  &  0.090  &  0.056  \\
24pt (12hr)     &  0.123  &  0.102  &  0.060  \\
36pt (18hr)     &  0.159  &  0.148  &  0.068  \\
\enddata

\end{deluxetable}

\begin{deluxetable}{lcccc}
\tablewidth{0pt}
\tabletypesize{\scriptsize}
\tablecolumns{5}
\tablecaption{\label{tab:logglimit}
Highest Surface Gravity Detectable for Different Shot Noise Levels}
\tablehead{ \colhead{$K_p$} &
            \colhead{Noise (ppt)} &
            \colhead{\logg$_{100\%}$} &
            \colhead{\logg$_{20\%}$} &
            \colhead{\logg$_{10\%}$}
          }
\startdata
    8.0 &  0.01316  &    4.49   &   5.31    &  5.79 \\
    8.5 &  0.01459  &    4.45   &   5.24    &  5.71 \\
    9.0 &  0.01676  &    4.39   &   5.16    &  5.61 \\
    9.5 &  0.01990  &    4.32   &   5.06    &  5.49 \\ 
   10.0 &  0.02428  &    4.23   &   4.96    &  5.36 \\
   10.5 &  0.03025  &    4.13   &   4.85    &  5.22 \\
   11.0 &  0.03815  &    4.03   &   4.74    &  5.09 \\
   11.5 &  0.04824  &    3.92   &   4.63    &  4.96 \\
   12.0 &  0.06083  &    3.81   &   4.53    &  4.84 \\
   12.5 &  0.07649  &    3.69   &   4.43    &  4.73 \\
   13.0 &  0.09649  &    3.56   &   4.33    &  4.63 \\
   13.5 &  0.12319  &    3.41   &   4.22    &  4.52 \\
   14.0 &  0.16199  &    3.24   &   4.10    &  4.40 \\
\enddata
\end{deluxetable}

\begin{deluxetable}{lccccccc}
\tablewidth{0pt}
\tabletypesize{\scriptsize}
\tablecolumns{8}
\tablecaption{\label{tab:f8}
Flicker Gravities for 27\,628 {\it Kepler} Stars}
\tablehead{ \colhead{{\it Kepler} ID} &
            \colhead{{\it Kepler} Magnitude} &
            \colhead{\fl\ \logg} &
            \colhead{\logg\ up error\tablenotemark{a}} &
            \colhead{\logg\ down error\tablenotemark{b}} &
            \colhead{Range\tablenotemark{c}} &
            \colhead{16pt RMS\tablenotemark{d}} &
            \colhead{\teff}
          }
\startdata
     1025494   &       11.822   &        3.874   &        0.100   &        0.100   &        0.266   &        0.055   &           6124  \\
     1026084   &       12.136   &        2.885   &        0.100   &        0.100   &        1.377   &        0.273   &           5039  \\
     1026255   &       12.509   &        3.640   &        0.100   &        0.100   &        1.971   &        0.361   &           7056  \\
     1026452   &       12.936   &        2.601   &        0.157   &        0.100   &        0.364   &        0.044   &           5087  \\
     1026475   &       11.872   &        3.948   &        0.100   &        0.100   &        0.457   &        0.055   &           6609  \\
     1026669   &       12.304   &        3.896   &        0.100   &        0.100   &        7.387   &        0.069   &           6303  \\
     1026861   &       11.001   &        3.763   &        0.100   &        0.100   &        0.320   &        0.063   &           7141  \\
     1026911   &       12.422   &        3.819   &        0.100   &        0.100   &        0.336   &        0.070   &           6594  \\
\enddata

\tablenotetext{a}{Error on \logg\ in the upward (main sequence) direction.}
\tablenotetext{b}{Error on \logg\ in the downward (red giant) direction.}
\tablenotetext{c}{Range value, corrected for {\it Kepler} magnitude, as described in \citet{bastien13}.}
\tablenotetext{d}{16pt RMS value used to calculate \fl.  See \citet{bastien13}}

\tablecomments{A portion of the table is shown for guidance regarding
content and format. The full table is available electronically.}

\end{deluxetable}

\end{document}